\documentclass[twocolumn,aps,prl,nofootinbib,superscriptaddress,tightenlines]{revtex4-1}

\usepackage{amsmath, amsfonts, amsthm, amssymb, graphicx, color}
\usepackage{hyperref}
\usepackage{subfigure}

\allowdisplaybreaks[3]

\def \bal#1\eal  {\begin{align} #1 \end{align}}
\newcommand{\be} {\begin{equation}}
\newcommand{\ee} {\end{equation}}
\newcommand{\nn} {\nonumber\\}

\newcommand{\ud} {\mathrm{d}}

\newcommand{\pd} {\partial}

\newcommand{\mc} {\mathcal}

\newcommand{\mn} {{\mu\nu}}

\newcommand{\ai}{{\alpha}}
\newcommand{\bi}{{\beta}}

\newcommand{\ri}{{\rho}}
\newcommand{\si}{{\sigma}}

\newcommand{\ep}{{\epsilon}}

\def\ba{\begin{eqnarray*}}
\def\ea{\end{eqnarray*}}

\def\stu{St\"uckelberg }

\def\d{\mathrm{d}}
\def\mn{_{\mu \nu}}

\def\({\left(}
\def\){\right)}

\def\mpl{M_{\rm Pl}}
\def\p{\partial}
\def\ie{{\em i.e. }}
\def\ien{{\em i.e.}}

\begin{document}

\title{Non-compact nonlinear sigma models}
\author{Claudia de Rham}
\email{claudia.deRham@case.edu}
\author{Andrew J.~Tolley}
\email{Andrew.J.Tolley@case.edu}
\author{Shuang-Yong Zhou}
\email{Shuangyong.Zhou@case.edu}
\affiliation{CERCA, Department of Physics, Case Western Reserve University, 10900 Euclid Ave, Cleveland, OH 44106, USA}
\date{\today}

\begin{abstract}

The target space of a nonlinear sigma model is usually required to be positive definite to avoid ghosts. We introduce a unique class of nonlinear sigma models where the target space metric has a Lorentzian signature, thus the associated group being non-compact. We show that the would-be ghost associated with the negative direction is fully projected out by 2 second-class constraints, and there exist stable solutions in this class of models. This result also has important implications for Lorentz--invariant massive gravity: There exist stable nontrivial vacua in massive gravity that are free from any linear vDVZ--discontinuity and a $\Lambda_2$ decoupling limit can be defined on these vacua.

\end{abstract}

\maketitle

\noindent{\bf Introduction and summary}
Nonlinear sigma models (NLSMs)~\cite{GellMann:1960np} are the umbrella name for many effective field theories~\cite{Weinberg:1978kz} from various areas of physics. See, e.g.,~\cite{nsmrev,nsmrev2,Zakrzewski:1989na} and references therein for a review. A NLSM maps from a base manifold (usually Minkowski) to a target space, a typical action of which is given by
\be
S_{\rm NLSM}  =  -\int\!\ud^D x \,\frac12\eta^{\mu\nu} \pd_\mu \phi^a \pd_\nu \phi^b f_{ab}(\phi)\,.
\ee
The $N$ dimensional target space metric $f_{ab}(\phi)$ is usually required to be positive definite to avoid ghost instabilities, that is, to avoid  having degrees of freedom (DoFs) with negative kinetic terms. In mathematical terms, the target space can only be Riemannian, but not pseudo-Riemannian. If the target space is symmetric, as usually considered, this requirement translates to the isometry group being compact, since a (real, semisimple) Lie algebra is compact if its Killing metric is sign definite.

However, there is an important well-known exception, which is the $p$-brane Nambu-Goto action from string/M theory~\cite{Becker:2007zj}
:
\be
S_{\rm NG} = - \frac{T_p}{2} \int \ud^{p+1} x  \sqrt{-\det\(\pd_\mu \phi^{a} \pd_\nu \phi^{b} f_{ab}(\phi)\)}\,,
\ee
where the target space metric $f_{ab}$ has a Lorentzian signature $(-,+...+)$. The reason why the $p$-brane action can avoid the ghost is because it has $p+1$ diffeomorphism invariances along the brane world-volume and the would-be ghost is merely a gauge DoF, projected out by a first class constraint. For example, the static gauge sets $\phi^a=x^a$, $a=0,...,p$, and the resulting target space becomes lower dimensional and manifestly positive definite. 

It is also possible to evade the compact group requirement by making use of  gauge fields, which has been utilized in supergravity model building~\cite{Cremmer:1979up, VanNieuwenhuizen:1981ae, nsmrev2}. A toy example is the $SU(1,1)/U(1)$ Cremmer--Julia NLSM~\cite{Cremmer:1979up}
\be
S_{\rm CJ} =\! \int \!\!\ud^D x \( | \pd_\mu \phi_1 -  A_\mu \phi_1 |^2 -  |\pd_\mu \phi_0-A_\mu \phi_0|^2 \),
\ee
where $A_\mu$ is an auxiliary $U(1)$ gauge field and $\phi_0$ and $\phi_1$ are constrained by $|\phi_0|^2-|\phi_1|^2=1$. (The associated gauge fields are auxiliary in the sense that we do not add a kinetic term for them.) Without the $U(1)$ gauge symmetry (\ien, $A_\mu=0$), this is a $SU(1,1)$ ``chiral'' NLSM and has a ghost. The introduction of the $U(1)$ gauge symmetry projects out this ghost, which can be seen explicitly by integrating out $A_\mu$, leading to  $S_{\rm CJ} = \int \ud^D x |\phi_0\pd_\mu\phi_1-\phi_1\pd_\mu\phi_0|^2$, whose target space is explicitly positive definite after gauge fixing and using the $SU(1,1)$ adjoint constraint. Generically, if one has a non-compact group $G$, the Killing metric will have both positive and negative signatures. To get a ghost-free NLSM, one may introduce a subgroup $H$ that gauges away all of the positive (or negative) directions of the Killing metric. 

To our knowledge, all the known evasions of the Riemannian/compact requirement make use of the auxiliary gauge trick, be it normal gauge symmetries or diffeomorphism invariances. Thus, all these evasions do not compromise the spirit of the Riemannian/compact requirement in the sense that first class constraints are manifestly employed to project out the would-be ghosts and after gauge fixing the target metric becomes explicitly positive definite.

In this letter, we point out that there is a unique way to evade the Riemannian/compact requirement of NLSMs by using second class constraints, thanks to a special square root and anti-symmetrization scheme. This novel class of NLSMs are given by
\bal
\label{genmasterlag}
S_{\rm mg\sigma} &=   \int\ud^Dx  \sum_{n=1}^{D} \bi_n(\phi) {X}^{\mu_1}_{[\mu_1}  {X}^{\mu_2}_{\mu_2} \cdots {X}^{\mu_n}_{\mu_n]} ,
\eal
where $\bi_n(\phi)$ are arbitrary functions of $\phi^a$, $f_{ab}(\phi)$ has a Lorentzian signature $(-,+...+)$ and for $N \ge D$
\be
{X}^\mu_{\, \nu} \! =\! \sqrt{\eta^{\mu\ri}\pd_\ri \phi^a \pd_\nu \phi^b f_{ab}(\phi)}, \,~ a=0,1,...,N-1\,.
\ee
It is understood that the principal branch of the metric square root is chosen. By examining perturbations on a general background, we will show that this class of NLSMs in general has $N-1$ physical DoFs and there exist solutions around which all the $N-1$ DoFs are free of ghost and gradient instabilities.

Incidentally, we note that the $p$-brane Nambu-Goto action can be re-casted as
\be
\label{NGX}
S_{\rm NG}  = -  \frac{T_p}{2}  \int \ud^{p+1} x X^{\mu_0}_{[\mu_0} X^{\mu_1}_{\mu_1}\cdots X^{\mu_p}_{\mu_p]} \,,
\ee
which is exactly the highest order term of Eq.~\eqref{genmasterlag} with $\bi_D=-T_p/2$. Also, the minimal model of (\ref{genmasterlag}), $\mc{L}_{\rm m} = - X^\mu_\mu$ with $N=D$, can be cast as $\mc{L}_m = - \frac12 \bar{\lambda}^{\ri\si} \pd_\ri \phi^\ai \pd_\si \phi^\bi \eta_{\ai\bi} - \frac12 \lambda_{\ai\bi}\eta^{\ai\bi}  =  - \frac12 {\eta}^{\ri\si} \pd_\ri \phi^\ai \pd_\si \phi^\bi \lambda_{\ai\bi} - \frac12 \eta_{\ai\bi} \bar{\lambda}^{\ai\bi}$, which strikingly resembles the Polyakov action. Despite these mathematical coincidences, however, there are significant physical differences. As mentioned above, $p+1$ DoFs of the $p$-brane action,  including the naive ghost, are projected out by $p+1$ first-class constraints, while for $S_{\rm mg\sigma}$, excluding the Nambu-Goto special case, only 1 DoF is projected out by 2 second-class constraints. We emphasize that the key for $S_{\rm mg\sigma}$ to have 2 second class constraints lies in its unique matrix square root and anti-symmetrization scheme. Geometrically, the Nambu-Goto action describes the Goldstone bosons nonlinearly realizing the diffeomorphism invariances spontaneously broken by placing the $p$-brane in the target space, while $S_{\rm mg\sigma}$ describes a special embedding of the flat spacetime into a fixed target space.

A reader familiar with massive gravity theories would have recognized that when $N=D$ with $f_{ab}=\eta_{ab}$ and $\bi_n$ are constant, 
$S_{\rm mg\sigma}$ is closely related to ghost-free massive gravity~\cite{deRham:2010ik,deRham:2010kj}:
\ba
\label{mastergravitylag}
S_{\rm mg} =\mpl^2\!\! \int\!\! \ud^{4} x \sqrt{-g} \left[ \frac{R}{2} \! +\!  m^2\!\sum_{n=1}^{4} \bi_n \mc{X}^{\mu_1}_{[\mu_1}   \cdots \mc{X}^{\mu_n}_{\mu_n]}  \right]\,,\nn
\ea
where $\mc{X}^\mu_\nu  \equiv \sqrt{g^{\mu\ri}\pd_\ri \phi^a \pd_\nu \phi^b \eta_{ab}}$. The reference metric $\eta_{ab}$ explicitly breaks diffeomorphism invariance. But, as with any gauge symmetry, it can be restored by introducing some extra fields $\phi^a$, called \stu fields, which encapsulate the vector (transverse) and scalar (longitudinal) modes of a massive graviton (at least in the decoupling limit\footnote{Of course the statement of who actually carries the DoFs is an ambiguous one in the full theory which is gauge invariant in the presence of the fields $\phi^a$. We also emphasize that a decoupling limit never changes the DoFs. Some DoF may decouple (which is the essence of a decoupling limit) but no DoF can ever ``appear" in such a limit. See Ref.~\cite{deRham:2014zqa} for more details.}). This is the unique, Poincar\'e invariant, ghost-free~\cite{deRham:2010ik,deRham:2010kj,Hassan:2011hr,Hassan:2011ea} massive gravity theory.

Very crudely speaking, $S_{\rm mg\sigma}$ arises by setting $g_{\mu\nu}= \eta_{\mu\nu}$ in $S_{\rm mg} $. However this is in general not a consistent thing to do, unless there exists a well-defined $\Lambda_2$ decoupling limit (see below) that decouples the extra DoFs inside the metric, leaving a decoupled sector which contains $S_{\rm mg\sigma}$. Phrasing it in terms of a decoupling limit is important, since it guarantees that all of the healthy properties of $S_{\rm mg} $ carry over to $S_{\rm mg\sigma}$.
The ghost-free validity of the extensions to the cases with $N\geq D$ with general $f_{ab}(\phi)$ and general $\beta_n(\phi)$ has been examined explicitly~\cite{Andrews:2013ora,deRham:2014lqa}.

The uniqueness of ghost-free massive gravity $S_{\rm mg}$~\cite{Matas:2015qxa} (essentially due to the uniqueness of the special square root and anti-symmetrization scheme of the potential) implies that the NLSM $S_{\rm mg\sigma}$, which we shall refer to as massive gravity NLSM, is the \emph{unique} NLSM whose target space can have one negative direction. Furthermore, since the unique matrix square root and anti-symmetrization scheme can only remove one (ghostly) DoF, a \emph{no-go theorem} may be stated that it is impossible to have a NLSM where there are two and more negative directions in the target space, without incorporating an auxiliary gauge procedure as that mentioned above (in which case the resulting target space becomes again positive definite after integration of those auxiliary variables and gauge fixing).

On the other hand, the study of the massive gravity NLSM $S_{\rm mg\sigma}$ alone has profound implications for ghost-free massive gravity. Since the Boulware-Deser ghost~\cite{Boulware:1973my} is eliminated in the action $S_{\rm mg}$, taking the $\Lambda_2$ decoupling limit $\mpl \rightarrow \infty$ {\it cannot} re-introduce the Boulware-Deser ghost. However, it is not known whether the $\Lambda_2$ decoupling limit will decouple any additional DoFs aside from the tensor modes from $S_{\rm mg\sigma}$. Alternatively, some of the existing DoFs may turn ghostly in the limit. In fact, there are arguments suggesting that these may happen. Ghost-free massive gravity is sometimes referred to as $\Lambda_3$ massive gravity, since around the trivial vacuum $g_{\mu\nu}=\eta_{\mu\nu},\phi^\ai =x^\ai$, the strong coupling scale is  $\Lambda_3=(\mpl m^2)^{{1/3}}$. Indeed, around this background, the longitudinal mode only acquires its kinetic term via mixing with the tensor modes. At the linear level, the longitudinal mode becomes a gauge DoF if the tensor modes are suppressed.
It is possible that this gauge symmetry can be maintained fully nonlinearly, which would imply that the $\Lambda_2$ decoupling limit is not a smooth limit to take.
This leads to a latent perception that $\Lambda_3$ is the highest strong coupling scale for healthy backgrounds in ghost-free massive gravity.

However our careful examination of the physical DoFs in $S_{\rm mg\sigma}$ shows that the gauge symmetry is {\it accidental} around the trivial vacuum and the $\Lambda_2$ decoupling limit is a smooth limit since all the vector and scalar modes can obtain healthy kinetic terms around some nontrivial vacuum
\be
\label{bg0}
g_{\mu\nu}=\eta_{\mu\nu}+\mc{O}(m^2), ~~\phi^a =\bar\phi^a(x) .
\ee
These type of vacua are very closely connected to those found in the cosmological setup~\cite{deRham:2010tw,D'Amico:2011jj}.
In unitary gauge $\phi^a=x^a$, this corresponds to having an almost flat vacuum $g_{\mu\nu}=\pd_\mu \bar\Phi^a \pd_\nu\bar\Phi^b \eta_{ab}+\mc{O}(m^2)$, where $\bar\Phi^a$ is the inverse or dual \stu fields $\bar\Phi^a(\bar \phi(x)) = x^a$. Although the metric is flat at leading order in $m^2$, it does not coincide with the reference Minkowski metric $\eta_{ab}$ due to the coordinate transformation between the two encoded in $\bar\Phi^a(x)$.

Around such a nontrivial vacuum, taking the limit $\mpl \rightarrow \infty$ forces the geometry of the $g_{\mu\nu}$ to be increasingly Ricci flat. For simplicity we take this Ricci flat metric to be Minkowski spacetime, and can then define the canonically normalized tensor modes $h_{\mu\nu}$ as
\be
g_{\mu\nu}=\eta_{\mu\nu}+\mpl^{-1}h_{\mu\nu}
\ee
so that on taking the full $\Lambda_2$ decoupling limit:
\be
\label{l2dl}
\mpl\to \infty,~~m\to 0,~~\Lambda_2=\sqrt{\mpl m}\to {\rm fixed},
\ee
ghost-free massive gravity $S_{\rm mg}$ reduces to
\be
S_{\rm mg}\! \to\!  \!\! \int\!\ud^4x \left[ -\frac{\mpl^2}4 h^{\mu\nu}\mc{E}_{\mu\nu}^{\ri\si}h_{\ri\si} \!+ \frac12 h_{\mu\nu}T^{\mu\nu} \right]+ \Lambda_2^4 \, S_{\rm mg\sigma} ,\nn
\ee
where $\mc{E}_{\mu\nu}^{\ri\si}$ is the linear Lichnerowicz operator and $T^{\mu\nu}$ is the energy momentum tensor from external matter fields.

Around these nontrivial vacua in the $\Lambda_2$ decoupling limit, only the tensor modes $h_{\mu\nu}$ couple to $T^{\mu\nu}$, so there is no linear  Van Dam--Veltman--Zakharov (vDVZ) discontinuity~\cite{vanDam:1970vg,Zakharov:1970cc}, and the phenomenological success of general relativity can be easily recovered. This is much like the well--known example of massive gravity around anti-de Sitter space (AdS), where there is no vDVZ discontinuity~\cite{Kogan:2000uy,Porrati:2000cp}. Also, analogous to the AdS case where the strong coupling scale is $\Lambda_2$ dressed by the AdS length scale, the strong coupling scale around these nontrivial vacua, $\Lambda_{2*}$, now depends on the specific background $\bar\phi^\ai$ chosen. Suppose a nontrivial vacuum has characteristic length $L$, then for $L>\Lambda_2^{-1}$ the strong coupling scale can be as high as $\Lambda_{2*} = \sqrt{\Lambda_2 L^{-1}}$~\cite{deRham:2016plk}.

Whilst more details, particularly the discussions on the $\Lambda_2$ decoupling limit and the nontrivial vacua of ghost-free massive gravity, will appear somewhere else~\cite{deRham:2016plk}, we present in the rest of the paper the DoF counting and the perturbative stability analysis on the massive gravity NLSM. A few extra points will also be discussed in the Discussions section.

\noindent{\bf Braneworld Bi-gravity interpretation} Before moving to an analysis of the NLSM, we give here a simple bi-gravity braneworld interpretation of the general NLSM with $N \ge D$ for generic target space metric $f_{ab}(Y)$, in the case where $\beta_n$ are constants. Consider $N$ dimensional Einstein gravity for a metric $f_{ab}$ with Planck mass $M_f$. Coordinates in this $N$ dimensional manifold are denoted by $Y^a$.
In this theory we consider a $D-1$ brane whose position is defined by the embedding $Y^a=\phi^a(x)$ where $x$ runs over $D$ dimensions. The induced metric on the brane is then given by
\be
f^I_{\mu\nu}(x) = f_{ab}(\phi) \partial_{\mu} \phi^a(x) \partial_{\nu} \phi^b(x) \, .
\ee
This metric describes the physical geometry of the brane and transforms in a standard way under brane diffeomorphisms and nonlinearly realized bulk diffeomorphisms. In addition, on the brane we add a new spin--2 DoF denoted by $g_{\mu\nu}(x)$. This DoF lives entirely on the brane and does not need to be defined through the bulk. Taken together, we may construct the brane tensor
\be
X^{\mu}_{\, \nu} = \sqrt{g^{\mu \rho} f^I_{\rho\nu}} \, ,
\ee
which transforms in a standard way under diffeomorphisms.
The full braneworld action may be taken to be
\bal
S_{\rm mg} &=\!\! \int\!\! \ud^{D} x \sqrt{-g} \left[ \frac{\mpl^2R[g]}{2} \! +\!  \Lambda_2^D\!\sum_{n=1}^{4} \bi_n {X}^{\mu_1}_{[\mu_1}   \cdots {X}^{\mu_n}_{\mu_n]}  \right]   \nn
& + \int \d^N Y \sqrt{-f} \frac{M_f^2 R[f]}{2} \, .
\eal
This is manifestly invariant under $N$ dimensional diffeomorphisms (nonlinearly realized on the brane). On taking the triple scaling limit
\be
M_f\to \infty, ~~\mpl\to \infty,~~m\to 0\,,~~\Lambda_2 \to {\rm fixed}\,,
\ee
and writing the metric $f_{ab}$ as
\be
f_{ab} = \bar f_{ab} + M_f^{-1} v_{ab} \, ,
\ee
the bulk gravitational modes $v_{ab}$ decouple, as do the fluctuations of $g$, leaving behind a massless $N$ dimensional graviton, a massless $D$ dimensional graviton and a decoupled $\Lambda_2$ NLSM. In this braneworld interpretation, unitary gauge $\phi^a(x) = x^a$ for $a=0, ..., D-1$ is simply the gauge in which the bulk coordinates are chosen to align with the brane.

\vskip 5pt
\noindent{\bf Perturbations on general backgrounds}
To count the number of DoFs of (\ref{genmasterlag}), one can perform a nonlinear Hamiltonian analysis~\cite{deRham:2016plk}. Equivalently, since linear perturbations on a generic background reflects exactly the same number of DoFs as in the full nonlinear Hamiltonian analysis, one can study linear perturbations on a generic background. More importantly, by studying linear perturbations, one can determine whether there exist non-trivial backgrounds where all of the DoFs are well behaved. For simplicity, we will focus on $N=D$ and $f_{ab}=\eta_{ab}$, which also has direct relevance to ghost-free massive gravity. As discussed above, the $\Lambda_2$ decoupling limit (\ref{l2dl}) would be well-defined if $\phi^a$ has $D-1$ well-behaved DoFs. Note that the BD ghost, eliminated in ghost-free massive gravity, can not re-emerge in a decoupling limit. 

For definiteness, we focus on the specific Lagrangian,
\be
\label{K2lag}
\mc{L}_{\ai_2} = 2  K^\mu_{[\mu} K^\nu_{\nu]} \,,~{\rm with}~K^\mu_\nu=\delta^\mu_\nu-X^\mu_\nu \,  .
\ee
Since the base and target space are now both Minkowski, we may identify the two and do not differentiate their indices carefully.
To analyze perturbations on a generic exact background of Eq.~(\ref{K2lag}) is technically involved, largely due to the complexity in dealing with the matrix square root. Here we will instead construct the background $\bar \phi^\mu$ itself perturbatively $\bar \phi^\mu=x^\mu + \ep B^\mu$ and then examine even smaller linear perturbations $V^\mu$ on this background:
\be
\label{phiexpBV}
\phi^\mu=\bar \phi^\mu(x)+ \varepsilon V^\mu .
\ee
The perturbations about the background $\bar \phi^\mu$ are encoded by the terms in $\mc{L}_{\ai_2}$ that are second order in $\varepsilon$. As we shall see below, it is sufficient for this analysis to expand the Lagrangian to second order in $\ep$. For simplicity, we shall work in $D=3$ dimensions when explicit calculations are needed.

The Lagrangian $\mc{L}_{\ai_2}$ is especially engineered so that one mode in $V^\mu$ is non-dynamical, which will manifest as a primary constraint when defining the conjugate momenta for the quadratic $\mc{L}_{\ai_2}$ to pass to the Hamiltonian formulation. However, this primary constraint takes a convoluted form, which hinders our further analysis. To circumvent this, we can make a linear field redefinition
\be
\label{VtoW}
V^0 = W^0 ,~~~~ V^i = W^i + T^i W^0    ,
\ee
such that $W^0$ is an auxiliary field, with $T^i$ determined by requiring that $\pd\mc{L}/\pd \dot{W}^0$ does not contain any $\dot{W}^\mu$. Then, defining the conjugate momenta $\pi_i $ for $W^i$, we obtain the quadratic Hamiltonian on the background $\bar\phi^\mu$:
\bal
\label{masHam}
\mc{H}_{\ai_2} &= \frac12 \pi_i \pi^i + \frac14 G_{ij}G^{ij}  + \mc{G}^{\ep}_1+ W^0 \left[ \pd_i \pi^i + \mc{G}^{\ep}_2  \right]
\nn
&~~~ - \ep^2 (W^0)^2  \mc{A}_{W^0} + \mc{O}(\ep^3)    ,
\eal
(up to boundary terms),
where we have defined $F^{\mu\nu} \equiv 2 \pd^{[\mu} B^{\nu]}, ~G^{\mu\nu} \equiv 2 \pd^{[\mu} W^{\nu]}$ and
\bal
\mc{A}_{W^0} &\equiv - \frac18 \big( \dot{F}_{ij}\dot{F}^{ij} + 2 \pd_k F_{ij} \pd^k F^{ij} + 2 \dot{F}^{0i}\pd^j F_{ij}
\nn
&~~~~~~~~~~~~ - \pd_i{F}^{0i}\pd_j {F}^{0j} -  \pd_k F_{0i} \pd^k F^{0i}  \big)  .
\eal
$\mc{G}^{\ep}_{1,2}\sim \mc{O}(\ep)$ do not depend on $W^0$, whose explicit forms are cumbersome to display here.

The $\mc{O}(\ep^0)$ of Eq.~(\ref{masHam}) is just Maxwell's theory, where the first class constraint $\pd_i \pi^i=0$ generates a gauge symmetry. If we had truncated up to $\mc{O}(\ep)$, $W^0$ would still be a Lagrange multiplier that enforces a modified first class constraint. When the $\mc{O}(\ep^2)$ corrections are included and the background $B^\mu$ is such that $\mc{A}_{W^0}\ne 0$, $W^0$ ceases to be a Lagrange multiplier. Indeed, we may integrate out $W^0$ and get
\be
\label{hamInt}
\mc{H}_{\ai_2} =   \frac12 \pi_i \pi^i + \frac14 G_{ij}G^{ij} + \mc{G}^{\ep}_1   \! +\! \frac{ \left( \pd_i \pi^i + \mc{G}^{\ep}_2   \right)^2}{4\ep^2 \mc{A}_{W^0}}   +  \mc{O}(\ep^3) .
\ee
We see that when $\mc{A}_{W^0}\neq 0$ all but one of the DoFs $\phi^\ai$ are activated. The reason why the leading term in Eq.~(\ref{hamInt}) is non-analytical in the limit $\ep\to 0$ is simply because our background itself is a perturbation around the trivial background $\bar \phi^\mu=x^\mu$, where a gauge symmetry emerges.

\vskip 5pt
\noindent{\bf Stability}
We now check whether there are backgrounds that are free of ghosts and gradient instabilities. Since the leading order of Eq.~(\ref{masHam}) is just the Maxwell theory, the transverse modes to leading order are just those of an Abelian gauge field, thus totally free of ghost or gradient instabilities. Therefore, we can focus on the longitudinal mode to determine the linear stability of the theory:
\be
\pi_i=\ep \frac{\pd_i }{\nabla^2} \chi ,~~~W_i=\frac{1}{\ep}\frac{\pd_i }{\sqrt{\nabla^2}} \psi,~~{\rm with}~\nabla^2=\pd_i \pd^i .
\ee
In what follows we assume that $B^\mu$ has a characteristic length scale $L$ and can be expanded in the form
\be
B^\mu = \left[ H^{\mu}_{\ri} x^\ri  +   \frac12 M^{\mu}{}_{\ri\si} \frac{ x^\ri x^\si}{L} + \mc{O}\left( x\left(\frac{x}{L}\right)^2 \right) \right]\,,
\ee
where $H$ and $M$ are constant in that expansion (in practise $L$ does not necessarily need to be a constant, it merely encapsulates the fact that $\p_\nu \bar \phi^\mu$ is not a constant matrix on distances larger than $L$). 
Making use of this local approximation and neglecting $\mc{O}\(\ep, {x}/{L} \)$ terms, the Hamiltonian for the longitudinal mode becomes
\bal
\label{longHfin}
\phantom{.}\hspace{-10pt} \mc{H}^{\parallel}_{\ai_2}\!\! &=\!   \frac{\tilde{\chi}^2}{4 \mc{A}_{W^0}}  \!+ \! \frac{1}{16} F_{ij}F^{ij} \pd_k\psi \pd^k \psi  \!+  \! F_{0i} \pd_j \psi  F_0{}^{[i}  \pd^{j]} \psi\,,\hspace{-10pt}
\eal
where $\tilde{\chi}\sim \chi + \pd (\pd B)^2\pd\psi+  (\pd B)^2\pd^2\psi$, $F_{\mu\nu}  = 2 H_{[\nu\mu]}$ and $\mc{A}_{W^0} =   \big(  M_{[0i]}{}^{i} M_{[0j]}{}^{j} + M_{[0i]j} M^{[0i]j} - M_{[ij]0} M^{[ij]}{}_{0} - 2 M_{[ij]k} M^{[ij]k} +2 M_{[0i]0} M^{[ij]}{}_{j} \big)/{2L^2}$.

Now, it is clear from Eq.~(\ref{longHfin}) that the gradient terms are manifestly (semi-)positive definite, so there is no gradient instability regardless of the values of $H\mn$.

On the other hand, $\mc{A}_{W^0}$ is not positive definite in general. But it is easy to find some explicit backgrounds that are stable. To see this, we note that up to $\mc{O}(\ep^2)$ the equation of motion for $B^\mu$ is just the Maxwell equation, which locally is given by $M_{[\mu\nu]}{}^\mu = 0$. Taking $M_{[\mu\nu]}{}^\mu = 0$ into account, one can show there are positive directions in the Hessian of $\mc{A}_{W^0}$ w.r.t.~$M_{\mu\ri\si}$, \ie there are background solutions where $\mc{A}_{W^0}$ is positive definite. For an explicit example in 3D, one may choose $M_{220}=1$ and $M_{\mu\ri\si}=0$ for all others, then we have $\mc{A}_{W^0} = 4>0$. Thus, at least locally within the characteristic length $L$, there are backgrounds that are free of ghosts and gradient instabilities.

Therefore, the $\Lambda_2$ decoupling limit (\ref{l2dl}) can be smoothly taken, and there are backgrounds where all the DoFs are healthy and the strong coupling scale may be pushed to $\Lambda_{2*}=\sqrt{\Lambda_2 L^{-1}}$~\cite{deRham:2016plk}.

\vskip 5pt
\noindent{\bf Discussions}
 In defining $S_{\rm mg\sigma}$, we have required $N\geq D$. The case $N<D$ has its own interest, and was for instance applied for the description of realistic condensed matter systems using the AdS/CFT correspondence in~\cite{Alberte:2014bua}. As shown in~\cite{Alberte:2013sma}, in the case where $N<D$ all the $N$ DoFs may propagate. By re-examining the Hamiltonian no-ghost proof of~\cite{Hassan:2011hr}, this can be shown to happen when the ``lapse'' squared of the reference metric, i.e., $-f_{00}+f_{0i}(f^{-1})^{ij}f_{0j}$, vanishes ($f_{ab}$ being zero-extended to be of the same dimension as $g_{\mu\nu}$). Whether or not those correspond to ghosts depend on the signature of the target space metric, and for a Lorentzian signature we expect a ghost-like DoF on arbitrary backgrounds.

It has been pointed out that for some parameter space ghost-free massive gravity does not exhibit the Vainshtein mechanism even for some seemingly benign initial-boundary conditions. We expect that for at least some of those cases, once the nontrival vector effects are carefully included, there will be no vDVZ discontinuity, and thus no need of a Vainshtein mechanism since the longitudinal mode decouples by itself. In other words, the non--trivial vacuum of the vectors are sufficient to implement the Vainshtein mechanism in a way which does not rely on external matter and does not affect the gravitational part of the theory. On these vacua, ghost-free massive gravity would then trivially pass most existing general relativity tests, while leading to a modification of gravity on distances larger than the graviton Compton wavelength.

For the $p$-brane NLSMs, the generalized Wess-Zumino-Novikov-Witten term~\cite{Wess:1971yu,Witten:1983tw,Novikov:1982ei} may be added:
\mbox{$S_{\rm wznw}\propto\int h_{a_0...a_p}(\phi)\ud\phi^{a_0}\wedge \cdots \wedge\ud\phi^{a_p}$}, where $h_{a_0...a_p}(\phi)$ is a $(p+1)$-form that depends on $\phi^a$, the existence of which implies the existence of the $p$-brane itself. In $S_{\rm mg\sigma}$, one may also consider this term. Since this term only contains one time derivative, its presence does not spoil the two second class constraints that project out the ghost.

\vskip 30pt
\noindent {\bf Acknowledgments:}
We would like to thank Paul Saffin for helpful discussions.
CdR is supported by a Department of Energy grant DE-SC0009946.
AJT and SYZ are supported by Department of Energy Early Career Award DE-SC0010600.

\bibliography{refs}

\end{document}